# The *w*-value: Market Implementation and γ-Core Stability in Pollution-Control Games

**Juanjuan Fan**[1], **Ying Wang**[2,*]

[a] School of Economics and Trade, Fujian Jiangxia University, Fuzhou, 350108, China

[b] School of Finance, Fujian Jiangxia University, Fuzhou, 350108, China

**Abstract** This paper proposes a novel solution concept, called the *w*-value, for cooperative pollution-control games. The *w*-value is axiomatically founded on three principles: Pareto Optimality (PO), Market Equilibrium (ME), and Fiscal Balance (FB), which together yield an efficient and implementable welfare distribution. We establish its existence and uniqueness under standard concavity and convexity assumptions, and provide a market mechanism that implements it through price constraints and an initial permit allocation. A key theoretical advantage is its strong stability: the *w*-value belongs to the γ-core, which makes it resistant to coalitional deviations. Moreover, the *w*-value can be obtained through a simple step-by-step procedure, which facilitates its application to complex real-world problems. A numerical illustration demonstrates the step-by-step calculation process and confirms the γ-core membership numerically. Finally, we apply the concept to international climate negotiations, showing how it provides a theoretical rationale for the shift from rigid mandatory emission reduction systems toward the flexible structure of Nationally Determined Contributions.

**Keywords:** *w*-value; Cooperative pollution-control game; γ-core; Pareto-optimality; Market-equilibrium; Fiscal-balance

* Corresponding author.
E-mail address: 2011008@fjjxu.edu.cn

## 1. Introduction

The pervasive challenge of environmental pollution stems fundamentally from the externalities associated with pollution emissions. Polluting agents often capture immediate, private benefits from their emissions, while the associated costs—increased pollution stocks and diminished welfare—are distributed across all agents, both present and future (Chander, 2017). In non-cooperative settings, individual agents, optimizing based solely on their private costs and benefits, inevitably generate aggregate emissions that exceed the socially efficient level. Achieving environmental efficiency typically necessitates cooperative agreements among agents, enabling the joint implementation of policies designed to internalize these external costs (Baumol and Oates, 1988). Within this context, cooperative game theory has emerged as a vital analytical framework for studying the formation, stability, and outcomes of such coalitions in pollution control (Chander and Tulkens, 1997).

Cooperative game theory, since its foundational work by von Neumann and Morgenstern, has developed along two primary, complementary pathways to derive solutions (Serrano, 2005). The first, often associated with the "Nash program", bridges cooperative and non-cooperative theory by modeling the negotiation process of cooperative games as a non-cooperative strategic interaction (Nash, 1953; Okada, 1996, 2012). The second employs an axiomatic approach, defining solutions based on a set of desirable properties such as efficiency, symmetry, and fairness. This approach has yielded seminal single-valued solution concepts like the Nash bargaining solution (Nash, 1950, 1953) and the Shapley value (Shapley, 1953), as well as set-valued solutions, most notably the core (Gillies, 1959). Various refinements of the core, including the α-core, β-core (Aumann, 1959), recursive core (Huang and Sjöström, 2003; Kóczy, 2007), Nash core (Okada, 2012), and γ-core (Chander and Tulkens, 1997), have been proposed, each based on different conjectures regarding the behavior of agents outside a forming coalition.

Despite this rich theoretical landscape, a significant gap persists between abstract solution concepts and their practical implementation in environmental policy. The γ-core, for instance, while theoretically appealing for games with positive externalities (Chander and Wooders, 2010), can be empty or multi-valued, limiting its prescriptive utility. More critically, much of the literature focuses on mathematical abstraction, often overlooking the essential role of specific, jointly implemented environmental policies required to sustain cooperation and internalize externalities in reality (Baumol and Oates, 1988). Achieving a Pareto-optimal outcome in pollution control universally requires cooperation, compelling agents to focus on two intertwined issues: first, identifying the specific policies that will lead to collective optimality, and second, determining a fair and stable distribution of the ensuing cooperation gains. The agents' primary concern is less about predicting complex retaliatory strategies of outsiders following a deviation, and more about the practical design of the cooperative regime itself.

This paper addresses this gap by analyzing the cooperative pollution control game through a policy-oriented lens. We demonstrate that from a mathematical standpoint, the game can be decomposed into two logically connected sub-games. The first (Sub-game 1) concerns the establishment of price constraints—effectively, policies like taxes or subsidies—on pollution emissions necessary to achieve Pareto optimality. The second (Sub-game 2) addresses the allocation of initial pollution emission permits among agents, which is fundamentally a distributional problem. We formally establish that a coalition, once formed, can be treated as a single, unified agent—a rationalizing assumption used in prior work (e.g., Chander, 2007; Chander and Wooders, 2010) but rigorously proven here.

Our analysis of Sub-game 1 confirms that a unique net price constraint is necessary to align individual incentives with the collective Pareto optimum, corroborating the classical finding of Baumol and Oates (1988). We further contribute a novel equilibrium condition: under this optimal constraint, the total subsidies for pollution abatement must equal the total taxes on emissions. This condition allows us to derive explicit optimal equilibrium prices. We prove that this framework incentivizes the formation of a grand coalition, ensuring the achievement of the pollution Pareto optimum.

The core distributive challenge, tackled in Sub-game 2, is the competition for initial emission permits. To resolve this, we introduce the concepts of w-initial pollution emission permits allocation ($w$-IPA), related Pareto optima, welfare distributions, and ultimately, the $w$-value. We demonstrate that under standard assumptions of strictly increasing and concave emission revenue functions and convex damage functions, the $w$-value provides a unique solution to the cooperative game. This solution, which is a subset of the γ-core, is rooted in market mechanisms, ensuring its practical implementability and making it a theoretically robust and appealing alternative to the γ-core for pollution control cooperation. Although developed in the context of mitigating negative externalities, the proposed framework is also applicable to cooperative games involving positive externalities, such as ecological conservation.

The remainder of this paper is structured as follows. Section 2 constructs the cooperative game model and formally decomposes it into Sub-game 1 and Sub-game 2. Section 3 analyzes the equilibrium of Sub-game 1, deriving the optimal price constraints. Section 4 investigates the equilibrium of Sub-game 2, establishing the $w$-value as the solution. Section 5 discusses the theoretical advantages of the $w$-value. A numerical illustration is provided in Section 6, and Section 7 concludes.

## 2. Public externality game model

### 2.1. Basic setting and notation

There are $I$ agents, indexed by $i$. The notation $-i$ denotes the subset of agents excluding agent $i$. Let $s_i$, $S_0$ and $S$ denote the amount of polluting input used by agent $i$, the initial stock of pollution, and the total pollution stock, respectively, such that $S=S_0+\sum_i s_i \geq 0$ ($\frac{\partial S}{\partial s_i}=1$). Similarly, let $\bar{s}_i$ and $\bar{s}=\sum_{i\in I}\bar{s}_i$ denote, respectively, the initial quantity of pollution-emission permits held by agent $i$ and the total initial quantity of pollution-emission permits available in the system..The price per unit of pollution emission (i.e., the price of pollution emission permits) is denoted by $t$. The marginal subsidy or tax assigned to agent $i$ for each unit of pollution stock exceeding a predefined threshold is denoted by $T_i$. The fixed lump-sum tax or subsidy imposed on agent $i$ is denoted by $\bar{t}_i$. Essentially, $t$ represents the price constraint faced by agents for "producing" a unit of pollution, while $T_i$ represents the marginal price constraint faced by agent $i$ for "consuming" a unit of pollution. Additionally, $\bar{t}_i$ is a fixed tax (or subsidy) imposed on agent $i$ , which does not vary with the "production" or "consumption" of pollution. Together, $t$, $T_i$, and $\bar{t}_i$ constitute all the price constraints imposed on agent $i$ regarding the "production" or "consumption" of pollution. Notably, $s_i$, $t$, $T_i$ and $\bar{t}_i$ are flow variables, while $S$ is a stock variable and $\bar{s}_i$ is an initial permit endowment. We assume that each unit of polluting input used emits one unit of pollution. Consequently, the variable $s_i$ also denotes the quantity of pollution emitted by agent $i$. The welfare of agent $i$ is specified as $w_i(s_i,S)=u_i(s_i)-d_i(S)$, where $u_i(s_i)$and $d_i(S)$ respectively represent the emission-related revenue function and pollution-damage function of agent $i$. Welfare is measured in monetary terms for all agents; hence, welfare is transferable among agents. That is, the game examined in this paper is a Transferable Utility (TU) game.

**Notation.** Throughout the paper, $s_i$ denotes the individual pollution emission of agent $i$, and $S = S_0 + \sum_{i=1}^{I} s_i$ denotes the aggregate pollution stock. The symbol $\overline{s}_i$ denotes the initial pollution-emission permit endowment of agent $i$, and $\overline{s} = \sum_{i=1}^{I} \overline{s}_i$ denotes the total initial permit endowment. For welfare, we distinguish between functions and allocated values. The lowercase $w_i(\cdot)$ denotes the welfare function of agent $i$. The uppercase $W_i$ denotes the welfare value allocated to agent $i$ under a given allocation or equilibrium. The uppercase $W^*$ denotes the total welfare at the Pareto optimum, i.e., $W^* = \sum_{i=1}^{I} w_i\,(s_i^*, S^*)$. For a coalition $C_{j,h}$ in coalition structure $h$, the uppercase $W_{j,h} = U_{j,h}(S_{j,h}) - D_{j,h}(S)$ denotes the total welfare of the coalition, where $U_{j,h}$ and $D_{j,h}$ are the coalition-level revenue and damage functions defined in Section 2.2.2. Finally, $s_i^*$ and $S^*$ denote Pareto-optimal equilibrium values, while $\tilde{s}_i$ and $\tilde{S}$ denote Nash equilibrium values under a specified coalition structure.

**Assumption 1.** Each agent acts to maximize its own welfare, and each coalition seeks to maximize the total welfare of all its members.

**Assumption 2.** Each emission revenue function $u_i(s_i)$ is strictly increasing, strictly concave, and satisfies the Inada condition $\lim_{s_i \to 0^+} u'_i(s_i) = +\infty$, while each damage function $d_i(S)$ is strictly increasing and strictly convex. That is,

$$u_i'(\mathrm{s_i}) > 0,\ u_i''(s_i) < 0,\ \lim_{s_i \to 0^+} u'_i(s_i) = +\infty,\ d_i'(S) > 0, \text{and}\ d_i''(S) > 0.$$

The Inada condition guarantees that the coalition's revenue-pooling problem (4) has an interior solution, $s_i > 0$ for all $i \in C_{j,h}$, for every $S_{j,h} > 0$.

**Assumption 3.** Each economic agent or coalition acts under the standard Nash conjecture; that is, it assumes that all other agents or coalitions will likewise act to maximize their own welfare.

Under these conditions, the solution to the maximization problem outlined below exists and is unique, as established in the literature (Chander, 2017). Within this framework, the core question this paper seeks to answer is: What constitutes the solution to the cooperative game of pollution control?

## 2.2. Basic model

### 2.2.1. Basic model of single agent

The welfare maximization problem for agent $i$ can be formulated as Model (1):

$$max\colon u_i(s_i) - d_i(S) + T_i\big(\textstyle\sum_{j=1}^{I} s_j - \sum_{j=1}^{I} \overline{s}_j\big) + t(\overline{s}_i - s_i) - \overline{t}_i \qquad (1)$$

In Model (1), agent $i$ chooses the emission $s_i$ to maximize its welfare, taking the price constraints $t$, $\bar{t}_i$, $T_i$ and the initial permit allocation $\bar{s}_i$ as given. The corresponding first-order condition ($2^c$) is derived, as presented in the Market equilibrium column of Table 1.

**Table 1**

Kuhn–Tucker conditions for optimality under pollution control

| Variable | Pareto optimality | Market equilibrium | Prices |
|---|---|---|---|
| $s_i$ | $\frac{\partial u_i}{\partial s_i} = \sum_{\bar{i}=1}^{I} \frac{\partial d_{\bar{i}}}{\partial S}$ ($2^o$) | $t - T_i = \frac{\partial u_i}{\partial s_i} - \frac{\partial d_i}{\partial S}$ ($2^c$) | $t - T_i = \sum_{\bar{i} \neq i} \frac{\partial d_{\bar{i}}}{\partial S}$ ($2^e$) |

Source: authors' compilation.

The optimal value of $s_i$ is determined through market transactions—that is, agent $i$ chooses $s_i$ optimally based on market prices $t$, $\bar{t}_i$ and $T_i$. In contrast, the optimal values of $\bar{s}_i$, $t$, $\bar{t}_i$ and $T_i$ are determined via a cooperative game. The essence of this cooperative game is to mathematically establish the equilibrium values of $\bar{s}_i$, $t$, $\bar{t}_i$ and $T_i$. Thus, from a mathematical perspective, the cooperative game can be decomposed into two interrelated sub-games: Sub-game 1 and Sub-game 2. Sub-game 1 involves determining the price constraints imposed on agents for "producing" or "consuming" pollution—specifically, it determines $t$, $\bar{t}_i$ and $T_i$. Sub-game 2 concerns the allocation of initial pollution emission permits among agents—that is, $\bar{s}_i$.

### 2.2.2. Basic model of coalition

The strategic behaviors of agents in the game can be classified as either non-cooperative or cooperative. Under a non-cooperative strategy, agents make decisions regarding pollution emissions or reductions based solely on their own welfare, without imposing constraints on their emission behavior. In essence, each agent acts independently and does not form any coalition with others. In contrast, a cooperative strategy entails that agents consider not only the effects of their emission-related actions on their own welfare but also the impacts on the welfare of others. In such cases, an agent adopting a cooperative strategy forms a coalition with at least one other agent.

If all agents adopt cooperative strategies and join a grand coalition, the outcome is a fully cooperative coalition structure. Otherwise, the outcome is an incomplete-cooperative coalition structure. Incomplete-cooperative coalition structures can be further divided into two subcategories: partially cooperative and fully non-cooperative. A fully non-cooperative coalition structure emerges when all agents employ non-cooperative strategies (i.e., every agent acts as a singleton), whereas any other incomplete-cooperative arrangement is termed a partially cooperative coalition structure.

There are $H$ possible coalition structures. Let $h$ index a coalition structure. In coalition structure $h$, all agents form $J_h$ coalitions, with $1 \leq J_h \leq I$. When $J_h = 1$, all agents adopt a cooperative strategy to form the grand coalition, corresponding to the fully cooperative coalition structure. When $J_h = I$, all agents adopt a non-cooperative strategy, resulting in the fully non-cooperative coalition structure. When $1 < J_h < I$, a partially cooperative coalition structure is formed. Together, the partially and fully non-cooperative structures comprise the set of incomplete-cooperative coalition structures.

Let $C$ denote a coalition and $I \setminus C$ denote its complement. Specifically, $C_{j,h}$ refers to the $j$-th coalition in coalition structure $h$. Thus, the coalition structure $h$ can be expressed as $P_h = \{C_{1,h}, C_{2,h}, \cdots, C_{J_h,h}\}$, which constitutes a partition of the set of all agents. Let $I_{j,h}$ represent the number of member agents in coalition $C_{j,h}$ within coalition structure $h$, indexed by $i_{j,h}$, such that $\sum_{j=1}^{J_h} I_{j,h} = I$ and $I_{j,h} \geq 1$. Denote by $s_{C_{j,h}} = (s_{i_{j,h}})_{i_{j,h} \in C_{j,h}}$ the joint strategy vector of coalition $C_{j,h}$. Denote by $S_{j,h} = \sum_{i_{j,h}=1}^{I_{j,h}} s_{i_{j,h}}$ and $\overline{s}_{j,h} = \sum_{i_{j,h}=1}^{I_{j,h}} \overline{s}_{i_{j,h}}$ the amount of polluting input used by coalition $C_{j,h}$ in coalition structure $h$, and the quantity of initial pollution emission permits held by coalition $C_{j,h}$ in coalition structure $h$, respectively. The welfare of coalition $C_{j,h}$ in coalition structure $h$ is defined as $W_{j,h} = U_{j,h}(S_{j,h}) - D_{j,h}(S)$, where the functions $U_{j,h}(S_{j,h}) = \sum_{i_{j,h}=1}^{I_{j,h}} u_{i_{j,h}}(s_{i_{j,h}})$ and $D_{j,h}(S) = \sum_{i_{j,h}=1}^{I_{j,h}} d_{i_{j,h}}(S)$ represent the emission revenue function and the damage function of coalition $C_{j,h}$ in coalition structure $h$, respectively.

We can formulate the welfare maximization problem of coalition $C_{j,h}$ in coalition structure $h$ as Model (3):

$$max\colon W_{j,h} = U_{j,h}(S_{j,h}) - D_{j,h}(S) = \sum_{i_{j,h}=1}^{I_{j,h}} u_{i_{j,h}}(s_{i_{j,h}}) - \sum_{i_{j,h}=1}^{I_{j,h}} d_{i_{j,h}}(S)$$

$$s.t.\ S_{j,h} = \sum_{i_{j,h}=1}^{I_{j,h}} s_{i_{j,h}},\ S = S_0 + \sum_{j=1}^{J_h} S_{j,h} \tag{3}$$

Model (3) maximizes the total welfare of all agents in coalition $C_{j,h}$ within coalition structure $h$, subject to the constraint that the availability of pollution emissions ($S_{j,h} = \sum_{i_{j,h}=1}^{I_{j,h}} s_{i_{j,h}}$) is satisfied. Because this constraint only defines the coalition's aggregate emission, the problem can be solved in two steps. First, for a given aggregate emission $S_{j,h}$, the coalition allocates emissions among its members so as to maximize total revenue, and the Lagrangian of this revenue-pooling problem is given by (4):

$$L_{j,h} = \sum_{i_{j,h}=1}^{I_{j,h}} u_{i_{j,h}}(s_{i_{j,h}}) - \lambda_{j,h}\left(\sum_{i_{j,h}=1}^{I_{j,h}} s_{i_{j,h}} - S_{j,h}\right) \tag{4}$$

The Greek letter $\lambda_{j,h}$ in (4) represents Lagrange multipliers, i.e., the shadow price of the coalition's aggregate emission $S_{j,h}$. Differentiating (4) with respect to $s_{i_{j,h}}$ yields $u'_{i_{j,h}}(s_{i_{j,h}}) = \lambda_{j,h}$, and the envelope theorem gives $U'_{j,h} = \lambda_{j,h}$. In the second step, the coalition chooses $S_{j,h}$ to maximize $U_{j,h}(S_{j,h}) - D_{j,h}(S)$, taking the emission levels of the other coalitions as given, so that the pollution stock $S$ increases one-for-one with $S_{j,h}$; the first-order condition is $U'_{j,h}(S_{j,h}) = D'_{j,h}(S)$, which determines $\lambda_{j,h}$. The equilibrium conditions for each agent $i_{j,h}$ are therefore derived as shown in (5):

$$u'_{i_{j,h}}(s_{i_{j,h}}) = \sum_{i_{j,h}=1}^{I_{j,h}} d'_{\bar{i}_{j,h}}(S) = \lambda_{j,h} \tag{5}$$

According to (5), in equilibrium, all members of the coalition share the same marginal condition: $\sum_{i_{j,h}=1}^{I_{j,h}} d'_{\bar{i}_{j,h}}(S)$. Therefore, we have:

**Lemma 1.** For any coalition that seeks to maximize the total welfare of its member agents, the following must hold:

$U'_{j,h} = \frac{\partial U_{j,h}}{\partial S_{j,h}} > 0,\ U''_{j,h} = \frac{\partial U'_{j,h}}{\partial S_{j,h}} < 0,\ D'_{j,h} = \frac{\partial D_{j,h}}{\partial S} > 0$ and $D''_{j,h} = \frac{\partial D'_{j,h}}{\partial S} > 0$.

**Proof.** See Appendix A. □

According to Lemma 1, once a coalition is formed, it acts as a single agent prior to its dissolution. Thus, in the subsequent analysis, any undissolved coalition is treated as an independent agent—i.e., such a coalition is regarded as a single agent. This premise is also adopted in some existing literature (see, e.g., Chander and Wooders, 2010; Chander, 2007); however, these studies have not mathematically established the rationality of this premise. Therefore, the aforementioned mathematical proof in this paper addresses the gap in the existing literature.

For the stability analysis below, we define the $\gamma$-deviation value and the $\gamma$-core explicitly.

**Definition 1. ($\gamma$-deviation value)** For any coalition $C \subseteq \{1,2,\cdots,I\}$, the $\gamma$-deviation value is

$$v_\gamma(C) = \max_{s_C} \sum_{i\in C} w_i(s_C, s_{-C}^{BR}(s_C)),$$

where $s_C = (s_i)_{i\in C}$ is the coalition's joint strategy, and $s_{-C}^{BR}(s_C)$ is the profile of Nash best responses of the outsiders $i \notin C$, each of whom maximizes its own welfare given $s_C$ and the strategies of the other outsiders. The coalition structure $\{C, \{i\}_{i\notin C}\}$, in which outsiders do not form a coalition and remain singletons, is called a $\gamma$-core coalition structure.

**Definition 2. ($\gamma$-core)** The $\gamma$-core of the game is the set of welfare allocations $(W_1, W_2, \cdots, W_I)$ such that $\sum_{i=1}^{I} W_i = W^*$, where $W^*$ is the total welfare at the Pareto optimum, and $\sum_{i\in C} W_i \geq v_\gamma(C)$ for every coalition $C \subseteq \{1,2,\cdots,I\}$.

## 3. Analysis of the equilibrium of Sub-game 1

**Lemma 2.** The final equilibrium of the cooperative pollution-control game must correspond to a Pareto optimum in pollution control.

**Proof.** See Appendix B. □

The Pareto-optimality problem can be formulated as Model (6):

$$max\colon \sum_{i=1}^{I} [u_i(s_i) - d_i(S)] \quad (6)$$

Model (6) maximizes the total welfare of all agents. The corresponding optimality conditions are provided in the Pareto optimality column of Table 1.

**Lemma 3.** The price conditions (2[e]) are necessary and sufficient to induce each agent to choose emission levels consistent with the Pareto optimum.

**Proof.** See Appendix C. □

Since the price conditions in (2[e]) are the necessary and sufficient conditions for the market equilibrium to achieve Pareto optimality, Eq. (2[e]) must be satisfied when calculating subsidies or taxes on pollution emissions, which is expressed as follows:

$$t - T_i = \sum_{\bar{i}\neq i} \frac{\partial d_{\bar{i}}}{\partial S} \quad (7)$$

Under Eq. (7), to attain Pareto optimal pollution emissions, each unit of emission reduction (or increase) should be subsidized such that the net subsidy equals the total marginal welfare effect of that unit on all other agents. If the total marginal welfare effect is positive, the net subsidy is positive; if negative, a tax should be imposed on the agent.

Furthermore, we can derive:

**Lemma 4.** In the process of implementing the rules specified in Lemma 3 to achieve Pareto optimality, the total amount of subsidies received by agents must equal the total amount of taxes paid by them.

**Proof.** See Appendix D. □

Economic Principle of Lemma 4: In pollution emission regulation, the subsidies received by agents for reducing emissions must originate entirely from the agents themselves, and the taxes or fees paid by agents for increasing emissions will ultimately be returned to them in full. That is, from the perspective of society as a whole, the fiscal revenues and expenditures arising from all agents' emission reductions and increases must necessarily balance. Lemma 4 can also be understood from another angle: if the government, acting as a non-profit public administrator, wishes to sustain the price constraint policy set forth in Lemma 3 over the long term, it must achieve fiscal balance. In fact, Lemma 3 only provides the price constraint conditions that should be imposed on emission-increasing (or emission-reducing) behaviors to internalize public externalities and achieve Pareto optimality. However, it does not specify the sources or uses of the funds required to implement these internalizing price constraints. Therefore, Lemma 3 alone is not a complete and sustainable policy. Only when Lemma 3 and Lemma 4 are combined do they constitute a complete and sustainable policy for achieving Pareto optimality.. Lemma 4 can be expressed as Eq. (8):

$$\sum_{i=1}^{I}[T_i(\sum_{j=1}^{I}s_j-\sum_{j=1}^{I}\bar{s}_j)+t(\bar{s}_i-s_i)-\bar{t}_i]=0 \quad (8)$$

Based on Eq. (7) and Eq. (8), the solutions to $t$, $\bar{t}_i$ and $T_i$ (with the derivation process detailed in Appendix E) are presented in Table 2:

**Table 2**
Solutions for $t$, $\bar{t}_i$ and $T_i$

| Type of price constraint | Value |
|---|---|
| Price Constraint I | $t=\sum_{\bar{i}=1}^{I}(\frac{\partial d_{\bar{i}}}{\partial S})$ (9)<br>$T_i=\frac{\partial d_i}{\partial S}$ (10)<br>$\sum_{i=1}^{I}s_i-\sum_{i=1}^{I}\bar{s}_i\neq 0$ (11)<br>$\bar{t}_i=0\ (\forall i)$ (12) |
| Price Constraint II | $t$, $\bar{t}_i$ and $T_i$ that satisfy Eqs. (7) and (8), excluding Price Constraint I |

In Table 2, $\frac{\partial d_i}{\partial S}$ denotes the marginal damage caused by a unit of pollution stock to agent $i$, and $\sum_{\bar{i}=1}^{I}\frac{\partial d_{\bar{i}}}{\partial S}$ represents the total marginal damage induced by a unit of pollution stock. We refer to the sets of $t$, $\bar{t}_i$ and $T_i$ that satisfy either Price Constraint I or Price Constraint II as optimal equilibrium prices.

From this, we deduce:

**Lemma 5.** For the market equilibrium to achieve Pareto optimality, the fiscal balance of public externality governance must be satisfied. This requires that $t$, $\bar{t}_i$ and $T_i$ correspond to the optimal equilibrium prices.

**Proof.** See Appendix E. □

If agents adopt a non-cooperative strategy, there are no constraints on pollution emissions in any agent; that is, $t = \bar{t}_i = T_i$. Then, from ($2^c$), we derive the equilibrium conditions under the no-emission-constraint scenario, expressed as (13):

$$\frac{\partial u_i(s_i)}{\partial s_i} = \frac{\partial d_i(S)}{\partial S} \tag{13}$$

By comparing Eq. ($2^o$) with Eq. (13), it can be observed that as long as at least one agent adopts the non-cooperative strategy, the game equilibrium fails to achieve the Pareto optimum in pollution control. Then, according to Lemma 2, we obtain the following result:

**Lemma 6.** In the final equilibrium of the cooperative game, all agents adopt cooperative strategies and form the grand coalition.

Note that Lemma 9 (presented below) also leads to the conclusion of Lemma 6. Several studies have reached similar conclusions to Lemma 6 (Chander, 2017; Maskin, 2003; De Clippel and Serrano, 2008, among others). It should be emphasized that, under Assumptions 2 and 3, the $\gamma$-core defined by Chander (2017; Chander and Wooders, 2010) is non-empty.

Moreover, according to the proof provided by Chander (2017; Chander and Wooders, 2010), Lemmas 5 and 6 can also be derived. The detailed reasoning is as follows. In the $\gamma$-core concept defined by Chander (2017; Chander and Wooders, 2010), each deviating coalition $C$ presumes that the resulting coalition structure after deviation is $\{C, \{i\}_{i \notin C}\}$, that is, players outside $C$ do not form a coalition and remain separate as singletons. We define this coalition structure as the $\gamma$-core coalition structure. The $\gamma$-core of the game is non-empty if there exists a welfare allocation $(W_1, W_2, \cdots, W_I)$ such that $\sum_{i=1}^{I} W_i = W^*$, where $W^* = \sum_{i=1}^{I} w_i(s_i^*, S^*)$, and, for each coalition $C \subseteq \{1,2,\cdots,I\}$, $\sum_{i \in C} W_i \geq v_\gamma(C)$. Hence $v_\gamma(C)$ is the $\gamma$-deviation value defined in Definition 1. Equivalently, $v_\gamma(C) = \sum_{i \in C} \widetilde{w}_i^\gamma(C)$, where $\widetilde{w}_i^\gamma(C)$ denotes the welfare of agent $i$ at the Nash equilibrium under the $\gamma$-core coalition structure. For all agents, welfare is measured in monetary terms; therefore, welfare is transferable among different agents. The total welfare at the Pareto optimum can be distributed among agents in any proportion, including the welfare allocation corresponding to the $\gamma$-core.

Based on the above analysis, the following proposition can be concluded.

**Proposition 1.** The equilibrium of Sub-game 1 is characterized as follows: To achieve the Pareto optimum of public-externality governance (which is the final equilibrium of the cooperative pollution-control game), all agents form a grand coalition to implement the price constraints $t$, $\bar{t}_i$ and $T_i$, where $t$, $\bar{t}_i$ and $T_i$ are identified as the optimal equilibrium prices.

## 4. Analysis of the equilibrium of Sub-game 2

As discussed in the Introduction, the core of Sub-game 2 is to determine the initial permit allocation. We first analyze the initial allocation based on the $w$-initial pollution-emission permit allocation ($w$-IPA) scheme, which is formally defined below.

### 4.1. Initial Permit Allocation and the $w$-IPA Scheme

First, the following result can be derived.

**Lemma 7.** Regardless of the initial allocation of pollution-emission permits, the final equilibrium total welfare of the pollution-control cooperative game remains unchanged, which equals the total welfare at the Pareto optimum. Pareto-optimal total welfare can be distributed among all agents by imposing Price Constraint I based on a specific initial allocation scheme of pollution-emission permits. Furthermore, the welfare of each agent increases with its initial number of pollution-emission permits and decreases with the total number of pollution-emission permits held by other agents.

**Proof.** See Appendix F. □

In essence, agents engaged in the cooperative pollution-control game aim to optimize the setting of $\bar{s}_i$, $t$, $T_i$, $\bar{t}_i$ to maximize their individual welfare. According to Proposition 1, the equilibrium solutions for $\bar{s}_i$, $t$, $T_i$, $\bar{t}_i$ are either Price Constraint I or Price Constraint II. Thus, Sub-game 2 corresponds to either the initial pollution-emission permit allocation game under Price Constraint I or that under Price Constraint II.

As established in Lemma 7, any Pareto-optimal total welfare distribution—including the welfare allocation achieved by implementing Price Constraint II based on a specific initial pollution-emission permit allocation scheme—can be realized by imposing Price Constraint I based on a distinct initial pollution-emission permit allocation scheme. Therefore, when analyzing Sub-game 2, it is only necessary to examine the initial pollution-emission permit allocation game under Price Constraint I. On the one hand, Price Constraint I specifies the rules for determining $\bar{s}_i$, $t$, $T_i$, $\bar{t}_i$. On the other hand, Lemma 7 indicates that each agent's welfare increases with its initial endowment of pollution emission permits and decreases with the total number of pollution-emission permits held by other agents. Consequently, the core issue for agents participating in the cooperative pollution-control game lies in competing for initial pollution-emission permits—which is precisely the problem to be addressed in Sub-game 2.

For further analysis, we introduce the following definitions.

**Definition 3.** The Nash equilibrium of pollution emissions under an incomplete-cooperative coalition structure serves as an initial pollution-emission permit allocation scheme. Based on this scheme, all agents jointly implement Price Constraint I to achieve the Pareto-optimal level of pollution emissions, thereby yielding a Category I welfare distribution scheme.

**Definition 4.** The Nash equilibrium pollution emissions resulting from a fully non-cooperative coalition structure constitute an initial pollution-emission permit allocation, referred to as the $w$-initial pollution-emission permit allocation ($w$-IPA).

**Definition 5.** On the basis of w-IPA, all agents jointly implement Price Constraint I to achieve the Pareto-optimal level of pollution emissions, which is referred to as the $w$-IPA Pareto-optimal level of pollution emissions.

**Definition 6 ($w$-value).** For each agent $i$, the $w$-value is defined as:

$$W_{i,w-value} = u_i(s_i^*) - d_i(S^*) + t^*(\tilde{s}_i - s_i^*) + T_i^*(\textstyle\sum_{j=1}^{I} s_j^* - \sum_{j=1}^{I} \bar{s}_j), \tag{14}$$

where $\tilde{s}_i$ is the Nash equilibrium emission of agent $i$ under the fully non-cooperative coalition structure (Definition 4), $(s_i^*, S^*)$ are the Pareto-optimal emissions and pollution stock, and $(t^*, T_i^*)$ are the optimal price constraints given by Price Constraint I (Table 2, Equations (9)–(10)). Equivalently, the $w$-value is the welfare of agent $i$ in the welfare distribution induced by the w-IPA scheme (Definitions 4–5). The economic interpretation of the $w$-value consists of three components: (i) the direct net benefit $u_i(s_i^*) - d_i(S^*)$ at the Pareto optimum; (ii) an emission adjustment transfer $t^*(\tilde{s}_i - s_i^*)$ that compensates the agent for deviations between its non-cooperative emission level and its Pareto-optimal emission level; and (iii) a fiscal redistribution term $T_i^*(\sum_{j=1}^{I} s_j^* - \sum_{j=1}^{I} \bar{s}_j)$, which redistributes the surplus from the price-constraint system.

According to the above definitions, the w-IPA welfare distribution constitutes a special Category I welfare distribution scheme. Specifically, the market mechanism in which all agents jointly enforce Price Constraint I to achieve the Pareto optimality of pollution emissions based on $w$-IPA not only enables the easy implementation of the $w$-value but also appears more operable than the market mechanism proposed by Serrano (1995).

**4.2. Characterization of the equilibrium of Sub-game 2**

Proposition 2 can be verified by successively proving the validity of the following five propositions (Propositions 3 – 7).

**Proposition 2.** Given that the final equilibrium of the cooperative pollution-control game is Pareto optimal, if $t$, $T_i$ and $\bar{t}_i$ satisfy Price Constraint I such that the market equilibrium achieves Pareto optimality, then each agent's final welfare must equal its .

**Proposition 3.** If agents fail to form a grand coalition (i.e., a fully cooperative coalition structure), the equilibrium of the pollution-control game can only be a Nash equilibrium under an incomplete-cooperative coalition structure.

**Proposition 4.** The equilibrium of the pollution-control game is attained only when all agents cooperate to form a grand coalition and achieve Pareto optimality. No Nash equilibrium under an incomplete-cooperative coalition structure can qualify as the game's equilibrium.

**Proposition 5.** Of the two welfare distribution schemes, the Category I welfare distribution scheme may serve as the equilibrium of the pollution-control game.

**Proposition 6.** Among all Category I welfare distribution schemes, only the $w$-IPA welfare distribution can serve as the equilibrium of the pollution-control game.

**Proposition 7.** None of the Category II welfare distribution schemes qualifies as the equilibrium of the pollution-control game.

Proposition 3 is self-evident and requires no further proof.

In an incomplete-cooperative coalition structure, the Nash equilibrium level of pollution emissions reflects the amount of pollution that agents independently choose to maximize their own welfare, as well as the amount they are both willing and capable of emitting. Moreover, if economic agents fail to form a fully cooperative coalition, the resulting coalition structure is inherently incomplete. Under such conditions, the only feasible pollution emissions correspond to the Nash equilibrium outcomes arising from non-cooperative behavior. Consequently, within an incomplete-cooperative coalition framework, the Nash equilibrium pollution emissions effectively define an initial allocation scheme for pollution permits—one that each agent is not only inclined to accept but is also bound to comply with.

Proposition 4 follows directly from Lemma 6 or from Lemma 9 below. Moreover, Lemma 9 further supports Proposition 5. However, to obtain Lemma 9, we must first prove Lemma 8.

**Lemma 8.** Under any given coalition structure, if a subset of agents adopts cooperative strategies, the total equilibrium pollution emissions will decrease relative to the fully non-cooperative structure. Specifically, for agents who adopt non-cooperative strategies, both their equilibrium emissions and welfare will increase; in contrast, for agents who adopt cooperative strategies, their equilibrium emissions will decrease, while the direction of changes in their welfare remains ambiguous.

**Proof.** See Appendix G. □

**Lemma 9.** For any incomplete coalition structure, let $\tilde{s} = (\tilde{s}_i)_{i\in\{1,2,\cdots,I\}}$ denote the Nash equilibrium emission profile. Set the initial permit allocation as $\bar{s}_i = \tilde{s}_i$ for all $i$, and implement Price Constraint I. Then:

(i) the resulting emission profile is Pareto optimal, i.e., $s_i = s_i^*$ for all $i \in \{1,2,\cdots,I\}$;

(ii) for every coalition $C \subseteq \{1,2,\cdots,I\}$, the total $w$-value allocated to members of $C$ is at least the $\gamma$-deviation value of $C$: $\sum_{i\in C} W_{i,w-value} \geq v_\gamma(C)$, where $v_\gamma(C)$ is defined in Definition 1.

**Proof.** See Appendix H. □

According to Lemma 9, under any incomplete-cooperative coalition structure, any agent (coalition) can achieve further welfare improvement in a Nash equilibrium. This implies that an incomplete-cooperative coalition structure will not be the final equilibrium of the cooperative game; only a fully cooperative coalition structure will serve as the final equilibrium of the game. Specifically, Lemma 9 can also yield the conclusion established in Lemma 6. Furthermore, Lemma 9 provides specific allocation schemes for pollution-emission permits under a fully cooperative coalition structure—namely, the initial allocation of pollution-emission permits derived from the Nash equilibrium pollution emissions of an incomplete-cooperative coalition structure. Based on this initial allocation, all agents collaborate to implement Price Constraint I, thereby achieving the Pareto optimality of pollution emissions and the corresponding welfare distribution. Notably, this welfare distribution falls under the Category I welfare distribution scheme described earlier. According to Lemma 9, under the Category I welfare distribution scheme, each coalition experiences improved welfare, thereby incentivizing coalitions to accept such improvements and ultimately leading to the conclusion presented in Proposition 5. In addition, part (ii) of Lemma 9 shows that the $w$-value allocation satisfies the coalitional stability condition $\sum_{i\in C} W_{i,w-value} \geq v_{\gamma}(C)$ for every coalition $C$, which is precisely the $\gamma$-core condition.

The conclusion of Proposition 6 follows from Lemma 10.

**Lemma 10.** Suppose all agents anticipate that the final equilibrium of the cooperative game of pollution control must correspond to the Pareto optimum of pollution control. To ensure the market equilibrium achieves this Pareto optimum, $t$, $T_i$ and $\bar{t}_i$ must satisfy the Price Constraint I such that among all Category I welfare distribution schemes, only the $w$-IPA welfare distribution can serve as the equilibrium of the cooperative pollution-control game.

**Proof.** See Appendix I. □

According to Lemma 10, among all Category I welfare distribution schemes, only the $w$-IPA welfare distribution can serve as the equilibrium of the game with public externalities, which constitutes the conclusion of Proposition 6.

The Pareto-optimal total welfare of the cooperative pollution-control game is constant. Accordingly, in any Category II welfare distribution scheme, at least one agent (say, Agent 1) has a welfare lower than that under the $w$-IPA welfare distribution. Under a Category II welfare distribution scheme, Agent 1 will reason as follows.

First, if he adopts a non-cooperative strategy to withdraw from the welfare distribution scheme, an incomplete-cooperative coalition structure will be formed. Second, according to Lemma 6 or Proposition 4, such an incomplete-cooperative coalition structure cannot be the final equilibrium of the cooperative game; instead, only the fully cooperative coalition structure (which corresponds to the Pareto optimum) will constitute the final equilibrium of the cooperative game. In other words, the final welfare allocation can only be a Category I welfare distribution scheme or a Category II welfare distribution scheme. Third, if a Category II welfare distribution scheme is formed again, there will be at least one other agent, say Agent 2, whose welfare is lower than that under the $w$-IPA welfare distribution, and this agent will replicate Agent 1's choice, thus causing the game to return to the first step (i.e., rendering the game a repeated game). Fourth, if a Category I welfare distribution scheme is formed, according to Proposition 6, of all Category I welfare distribution schemes, only the $w$-IPA welfare distribution can serve as the equilibrium of the game with public externalities, meaning the final welfare of each agent must equal its $w$-value: $W_i = W_{i,w-value}$. This establishes Proposition 2.

According to Proposition 2, the final welfare of each agent must equal its $w$-value under the fully cooperative coalition structure. However, Proposition 2 alone does not address the stability of this allocation against coalitional deviations. In what follows, we formally establish that the $w$-value allocation is resilient to such deviations by proving its membership in the $\gamma$-core. This requires, first and foremost, a precise definition of the $\gamma$-deviation value—provided in Definition 1 and used throughout the stability analysis below.

## 5. Discussion

We propose that the $w$-value constitutes a theoretically more robust solution concept for the cooperative pollution-control game, under the following assumptions: (i) all agents seek to maximize their own welfare; (ii) the emission revenue functions are strictly increasing and strictly concave, while the damage functions are strictly increasing and strictly convex.

The $w$-value allocation is contained in the $\gamma$-core.

Recall from Definition 1 that the $\gamma$-core coalition structure is $\{C, \{i\}_{i \notin C}\}$, in which players outside $C$ do not form a coalition and remain separate as singletons. For the same coalition $C$, the $\gamma$-deviation value is

$$v_\gamma(C) = \max_{s_C} \sum_{i \in C} w_i(s_C, s_{-C}^{BR}(s_C)),$$

where $s_C = (s_i)_{i \in C}$ is the coalition's joint strategy, and $s_{-C}^{BR}(s_C)$ is the profile of Nash best responses of the outsiders $i \notin C$.

On the basis of the $\gamma$-core coalition structure, Chander (2007) and Chander and Wooders (2010) defined the $\gamma$-core.

Recall from Definition 2 that the $\gamma$-core of the game is a welfare allocation $(W_1, W_2, \cdots, W_I)$ such that $\sum_{i=1}^{I} W_i = W^*$, where $W^*$ is the total welfare at the Pareto optimum, and $\sum_{i\in C} W_i \geq v_\gamma(C)$ for every coalition $C \subseteq \{1,2,3,\cdots,I\}$.

Let $\widetilde{w}_i^\gamma(C)$ denote the welfare of agent $i$ at the Nash equilibrium resulting from the $\gamma$-core coalition structure. Equivalently, the $\gamma$-deviation value can be written as $v_\gamma(C) = \sum_{i\in C} \widetilde{w}_i^\gamma(C)$.

First, by the definition of the $w$-value, the allocation $(W_{1,w-value}, W_{2,\, w-value}, \cdots, W_{I,w-value})$ satisfies $\sum_{i=1}^{I} W_{i,w-value} = W^*$, which is the efficiency condition of the $\gamma$-core. Second, by Lemma 9(ii), for every coalition $C \subseteq \{1,2,\cdots,I\}$, $\sum_{i\in C} W_{i,w-value} \geq v_\gamma(C)$. Therefore, the *w*-value allocation satisfies both conditions of the $\gamma$-core. Hence, the *w*-value is contained in the $\gamma$-core. Furthermore, the *w-value* incorporates a market-based mechanism that ensures the existence and uniqueness of the market equilibrium that achieves Pareto optimality, making it a theoretically more compelling solution concept than the $\gamma$-core for the cooperative pollution-control game.

On the Choice of the $\gamma$-core. Regarding the choice of stability concepts, this paper adopts the $\gamma$-core rather than the recursive core (Kóczy, 2007) or the sustainable core (Habis and Csercsik, 2014) for the following reasons. First, the $\gamma$-core assumes that outsiders to a deviating coalition remain as singletons, which closely matches the reality of international environmental agreements characterized by "voluntary participation and free exit." Second, Chander and Tulkens (1997) have proven the non-emptiness of the $\gamma$-core economies with multilateral environmental externalities, providing a solid theoretical foundation for our analysis. Third, the recursive core requires recursive solutions to residual games, which entails prohibitive computational complexity under our continuous strategy space. Fourth, the sustainable core is primarily designed for long-term dynamic games, whereas this paper focuses on equilibrium analysis of static games. Therefore, the $\gamma$-core is the most appropriate stability concept under our framework.

Although this paper focuses on the cooperative pollution-control game, our findings can be extended to other cooperative games involving negative-externality control, as well as to games concerning positive-externality protection, such as the preservation of beneficial ecological environments. Let $G$ denote the final stock of a good exhibiting positive externality, and let $g_i$ represent the amount of the good produced by agent $i$. The functions $r_i(G)$ and $f_i(g_i)$ denote the revenue function and production function of agent $i$, respectively. Similarly, it can be shown that the

*w*-value also serves as a solution concept for cooperative games of positive externality protection, under the assumptions that (i) all agents maximize their own welfare, and (ii) each revenue function $r_i(G)$ is strictly increasing and strictly concave, while each production function $f_i(g_i)$ is strictly increasing and strictly convex—that is, $r'_i(G) > 0$, $r''_i(G) < 0$, $f'_i(g_i) > 0$, $f''_i(g_i) > 0$.

# 6. An illustrative example

## 6.1 Calculation steps of *w*-value

The calculation steps of the *w*-value can be outlined as follows:

(1) Calculate the Nash equilibrium pollution emissions of each agent, denoted by $\tilde{s}_i$, under the fully non-cooperative coalition structure according to Equation (13), and designate them as the initial pollution emission allowances of each agent, i.e., set $\bar{s}_i = \tilde{s}_i$.

(2) Calculate the Pareto-optimal pollution emissions of each agent, denoted by $s_i^*$, according to Equation (2°) in Table 1.

(3) Substitute the values of $\bar{s}_i = \tilde{s}_i$, $s_i^*$, and $S^*$ into Equations (9) and (10), and calculate the price constraint $t^*$ and $T_i^*$.

(4) Substitute the values of $\bar{s}_i = \tilde{s}_i$, $s_i^*$, $S^*$, $t^*$, and $T_i^*$ into Equation (1), and calculate the *w*-value: $W_{i,w-value} = u_i(s_i^*) - d_i(S^*) + T_i^*(\sum_{j=1}^{I} s_j^* - \sum_{j=1}^{I} \tilde{s}_j) + t^*(\tilde{s}_i - s_i^*)$

## 6.2 An illustrative example of *w*-value

Consider a cooperative pollution-control game with $I = 4$. The revenue functions and damage functions are specified in Table 3.

Table 3

Revenue functions and damage functions

| Agent $i$ | Revenue function $u_i(s_i)$ | Marginal revenue $u'_i(s_i)$ | Damage function $d_i(S)$ | Marginal damage $d'_i(S)$ |
|---|---|---|---|---|
| 1 | $u_1(s_1) = 10 + 6s_1^{1/2}$ | $3s_1^{-1/2}$ | $d_1(S) = S^2$ | $2S$ |
| 2 | $u_2(s_2) = 6 + 2s_2^{1/2}$ | $s_2^{-1/2}$ | $d_2(S) = 2S^2$ | $4S$ |
| 3 | $u_3(s_3) = 5 + 3s_3^{1/3}$ | $s_3^{-2/3}$ | $d_3(S) = 3S^2$ | $6S$ |
| 4 | $u_4(s_4) = 8 + 4s_4^{1/2}$ | $2s_4^{-1/2}$ | $d_4(S) = S^3$ | $3S^2$ |

The initial pollution stock is $S_0$=0, so the total pollution stock is $S$=$\sum_{i=1}^{4} s_i$.

Under the fully non-cooperative coalition structure, the Nash equilibrium emissions are

$$\tilde{s} = (\tilde{s}_1, \tilde{s}_2, \tilde{s}_3, \tilde{s}_4) = (1.182, 0.033, 0.042, 0.123),$$

with total emission $\sum_{i=1}^{4} \tilde{s}_i = 1.380$ and Nash equilibrium welfare

$$(\tilde{w}_1, \tilde{w}_2, \tilde{w}_3, \tilde{w}_4) = (14.620, 2.556, 0.333, 6.775),\ \sum_{i=1}^{4} \tilde{w}_i = 24.284.$$

The Pareto-optimal emissions are $s^* = (s_1^*, s_2^*, s_3^*, s_4^*) = (0.247, 0.027, 0.067, 0.110)$, with total pollution stock $S^* = \sum_{i=1}^{4} s_i^* = 0.452$. The optimal price constraints under Price Constraint I are

$$t^* = D'(S^*) = 12S^* + 3(S^*)^2 \approx 6.037,$$

$$T^* = (T_1^*, T_2^*, T_3^*, T_4^*) = (0.904, 1.808, 2.712, 0.613).$$

The $w$-value of each agent is

$$W_{1,w-value} = 17.583,\quad W_{2,w-value} = 4.279,\quad W_{3,w-value} = 2.940,\quad W_{4,w-value} = 8.743,$$

with total

$$\sum_{i=1}^{4} W_{i,w-value} = 33.545.$$

Compared with the fully non-cooperative Nash equilibrium, the welfare improvements are

$(2.963, 1.723, 2.607, 1.968)$, $\sum_{i=1}^{4} (W_{i,w-value} - \tilde{w}_i) = 9.261.$

**Table 4**
Result of the illustrative example

| Agent $i$ | $w$-value $W_{i,w-value}$ | Nash welfare $\tilde{w}_i$ | Welfare improvement |
|---|---|---|---|
| 1 | 17.583 | 14.620 | 2.963 |
| 2 | 4.279 | 2.556 | 1.723 |
| 3 | 2.940 | 0.333 | 2.607 |
| 4 | 8.743 | 6.775 | 1.968 |
| Sum | 33.545 | 24.284 | 9.261 |

Table 5 compares the $w$-value allocation with the $\gamma$-deviation values of all non-singleton deviating coalitions. For every coalition $C \subseteq \{1,2,3,4\}$, the total $w$-value allocated to the members of $C$ is at least the $\gamma$-deviation value $v_\gamma(C)$: $\sum_{i \in C} W_{i,w-value} \geq v_\gamma(C)$. This numerically confirms that the $w$-value allocation belongs to the $\gamma$-core.

Table 5
Comparison between w-value and $\gamma$-core

| Coalition $C$ | $\gamma$-deviation value $v_\gamma(C)$ | $w$-value of coalition $C$, $\sum_{i \in C} W_{i,w-value}$ | Difference |
|---|---|---|---|
| {1,2} | 16.854 | 21.862 | 5.008 |
| {1,3} | 15.189 | 20.523 | 5.334 |
| {1,4} | 22.672 | 26.326 | 3.654 |
| {2,3} | 2.661 | 7.219 | 4.558 |
| {2,4} | 8.932 | 13.022 | 4.090 |
| {3,4} | 6.668 | 11.683 | 5.015 |
| {1,2,3} | 19.228 | 24.802 | 5.574 |
| {1,2,4} | 27.857 | 30.605 | 2.748 |
| {1,3,4} | 26.907 | 29.266 | 2.359 |
| {2,3,4} | 9.238 | 15.962 | 6.724 |
| {1,2,3,4} | 33.545 | 33.545 | 0 |

Table 6 presents the equilibrium pollution emissions and welfare of all agents under all coalition structures. It illustrates Lemma 8: under any incomplete-cooperative structure, if a subset of agents adopts cooperative strategies, total emissions decrease. Agents adopting non-cooperative strategies increase both their emissions and welfare, whereas agents adopting cooperative strategies reduce their emissions but experience ambiguous welfare changes. Notably, the fully cooperative structure in the last row corresponds to the Pareto optimum.

Table 7 compares the initial equilibrium welfare under each coalition structure with the welfare achieved under the w-IPA scheme (i.e., implementing Price Constraint I based on the Nash equilibrium permit allocation). For any incomplete-cooperative coalition structure, this mechanism leads to Pareto optimality and strictly improves total welfare, which is consistent with Lemma 9. For the fully cooperative structure, the initial equilibrium is already the Pareto optimum, and the welfare change is zero (the reported 0.001 difference is due to rounding).

In summary, the numerical example illustrates three main points. First, the *w*-value allocation is well defined and yields a strict welfare improvement for every agent relative to the fully non-cooperative Nash equilibrium (as shown in Table 4). Second, the *w*-value allocation belongs to the $\gamma$-core: for every deviating coalition $C$, the total *w*-value allocated to its members is at least the $\gamma$-deviation value $v_\gamma(C)$ (as shown in Table 5). Third, under any incomplete-cooperative coalition structure, implementing Price Constraint I based on the Nash equilibrium permit allocation leads to Pareto optimality, as stated in Lemma 9, and improves every agent's welfare (as shown in Tables 6 and 7).

**Table 6**
Equilibrium emissions and welfare under all coalition structures

| Serial number | Coalition structures | $s_1$ | $s_2$ | $s_3$ | $s_4$ | Total emission | $w_1$ | $w_2$ | $w_3$ | $w_4$ | Total welfare |
|---|---|---|---|---|---|---|---|---|---|---|---|
| 1 | [1], [2], [3], [4] | 1.182 | 0.033 | 0.042 | 0.123 | 1.380 | 14.620 | 2.556 | 0.333 | 6.775 | 24.284 |
| 2 | [1], [2], [3, 4] | 1.259 | 0.035 | 0.020 | 0.022 | 1.337 | 14.945 | 2.800 | 0.459 | 6.209 | 24.413 |
| 3 | [1], [2, 3], [4] | 1.210 | 0.005 | 0.020 | 0.129 | 1.364 | 14.740 | 2.427 | 0.234 | 6.898 | 24.299 |
| 4 | [1], [2, 4], [3] | 1.253 | 0.009 | 0.044 | 0.035 | 1.340 | 14.920 | 2.594 | 0.671 | 6.338 | 24.523 |
| 5 | [1], [2, 3, 4] | 1.292 | 0.003 | 0.013 | 0.012 | 1.320 | 15.079 | 2.626 | 0.475 | 6.137 | 24.317 |
| 6 | [1, 2], [3], [4] | 0.280 | 0.031 | 0.074 | 0.559 | 0.944 | 12.285 | 4.569 | 3.585 | 10.148 | 30.587 |
| 7 | [1, 2], [3, 4] | 0.481 | 0.053 | 0.070 | 0.116 | 0.721 | 13.644 | 5.424 | 4.679 | 8.986 | 32.733 |
| 8 | [1, 3], [2], [4] | 0.167 | 0.074 | 0.050 | 0.626 | 0.918 | 11.609 | 4.860 | 3.580 | 10.393 | 30.442 |
| 9 | [1, 3], [2, 4] | 0.304 | 0.059 | 0.079 | 0.237 | 0.680 | 12.848 | 5.563 | 4.900 | 9.634 | 32.945 |
| 10 | [1, 4], [2], [3] | 0.511 | 0.078 | 0.080 | 0.227 | 0.896 | 13.485 | 4.954 | 3.887 | 9.187 | 31.513 |
| 11 | [1, 4], [2, 3] | 0.564 | 0.013 | 0.039 | 0.251 | 0.868 | 13.755 | 4.725 | 3.760 | 9.350 | 31.590 |
| 12 | [1, 2, 3], [4] | 0.082 | 0.009 | 0.029 | 0.756 | 0.876 | 10.946 | 4.657 | 3.625 | 10.806 | 30.034 |
| 13 | [1, 2, 4], [3] | 0.335 | 0.037 | 0.129 | 0.149 | 0.651 | 13.051 | 5.538 | 5.245 | 9.268 | 33.102 |
| 14 | [1, 3, 4], [2] | 0.254 | 0.170 | 0.069 | 0.113 | 0.606 | 12.657 | 6.090 | 5.128 | 9.122 | 32.997 |
| 15 | [1, 2, 3, 4] | 0.247 | 0.027 | 0.067 | 0.110 | 0.452 | 12.778 | 5.923 | 5.609 | 9.233 | 33.544 |

Note: For the fully non-cooperative structure [1], [2], [3], [4], the equilibrium is the Nash equilibrium. For the fully cooperative structure [1, 2, 3, 4], the equilibrium corresponds to the Pareto optimum $(s_i^*, S^*)$.

**Table 7**
Welfare comparison between initial coalition structures and the w-IPA scheme

| Serial number | Coalition structure | Initial equilibrium welfare | Welfare under the w-IPA scheme | Welfare improvement |
|---|---|---|---|---|
| 1 | [1], [2], [3], [4] | 24.284 | 33.545 | 9.261 |
| 2 | [1], [2], [3, 4] | 24.413 | 33.545 | 9.132 |
| 3 | [1], [2, 3], [4] | 24.299 | 33.545 | 9.246 |
| 4 | [1], [2, 4], [3] | 24.523 | 33.545 | 9.022 |
| 5 | [1], [2, 3, 4] | 24.317 | 33.545 | 9.228 |
| 6 | [1, 2], [3], [4] | 30.587 | 33.545 | 2.958 |
| 7 | [1, 2], [3, 4] | 32.733 | 33.545 | 0.812 |
| 8 | [1, 3], [2], [4] | 30.442 | 33.545 | 3.103 |
| 9 | [1, 3], [2, 4] | 32.945 | 33.545 | 0.600 |
| 10 | [1, 4], [2], [3] | 31.513 | 33.545 | 2.032 |
| 11 | [1, 4], [2, 3] | 31.590 | 33.545 | 1.955 |
| 12 | [1, 2, 3], [4] | 30.034 | 33.545 | 3.511 |
| 13 | [1, 2, 4], [3] | 33.102 | 33.545 | 0.443 |
| 14 | [1, 3, 4], [2] | 32.997 | 33.545 | 0.548 |
| 15 | [1, 2, 3, 4] | 33.544 | 33.545 | 0.001 |

Note: For any incomplete-cooperative coalition structure, implementing Price Constraint I based on the Nash equilibrium permit allocation (the w-IPA scheme) leads to the Pareto optimum and improves total welfare, consistent with Lemma 9. For the fully cooperative structure in row 15, the initial equilibrium is already the Pareto optimum; the difference of 0.001 is due to rounding.

## 7. Conclusion

This paper introduces a new solution concept, referred to as the $w$-value, for cooperative pollution-control games. By decomposing the game into two logically connected sub-games, we demonstrate that the *w-value* arises from a practical market mechanism: Sub-game 1 derives the optimal price constraints (Price Constraint I) to internalize externalities, and Sub-game 2 determines the initial permit allocation based on the fully non-cooperative Nash equilibrium ($w$-IPA). We formally define the $\gamma$-deviation value $v_\gamma(C)$ and show that, for any deviating coalition $C$, the total *w-value* allocated to its members is at least $v_\gamma(C)$.Consequently, the *w-value* allocation is contained in the $\gamma$-core. Unlike the abstract $\gamma$-core, the *w-value* incorporates a market-based mechanism that ensures Pareto optimality and provides a readily implementable welfare distribution, rendering it a theoretically robust and practically operable solution concept for pollution control.

Furthermore, the $w$-value is also a solution to the cooperative game of positive-externality protection underanalogous assumptions: all agents maximize their own welfare; revenue functions $r_i(G)$ are strictly increasing and strictly concave; and production functions $f_i(g_i)$ are strictly increasing and strictly convex (i.e., $r'_i(G) > 0$, $r''_i(G) < 0$, $f'_i(g_i) > 0$, and $f''_i(g_i) > 0$). This paper establishes a theoretical framework for cooperative games involving the control of negative externalities or the protection of positive externalities. Future research could relax these strict concavity and convexity assumptions to further sharpen the game-theoretic analysis.

## CRediT authorship contribution statement

**Juanjuan Fan:** Conceptualization, Formal analysis, Writing – original draft. **Ying Wang:** Data curation, Visualization, Writing – original draft (appendix assistance). **All authors:** Writing – review and editing.

## Data availability

No data was used for the research described in the article.

## Statements and Declarations

We have no conflicts of interest.

## Acknowledgements

This work was supported by Fujian Research Center for Xi Jinping Thought on Socialism with Chinese Characteristics for a New Era (Grant No. FJ2023XZB054).

## Compliance with Ethical Standards

Conflict of Interest The authors declare competing interests.

# Appendix A. Proof of Lemma 1

By Assumption 2, $d_i(S)$ is strictly increasing and strictly convex for all $i$. The coalition damage function is defined as $D_{j,h}(S) = \sum_{i \in C_{j,h}} d_i(S)$. Taking the first and second derivatives with respect to $S$ yields:

$$D'_{j,h}(S) = \sum_{i \in C_{j,h}} d'_i(S) > 0$$

and

$$D''_{j,h}(S) = \sum_{i \in C_{j,h}} d''_i(S) > 0.$$

Next, consider the coalition revenue function $U_{j,h}(S_{j,h})$. When the coalition acts as a single agent to maximize its total revenue subject to a given aggregate emission $S_{j,h}$, it solves the following optimization problem:

$$U_{j,h}(S_{j,h}) = \max_{(s_i)_{i \in C_{j,h}}} \sum_{i \in C_{j,h}} u_i(s_i) \text{ subject to } \sum_{i \in C_{j,h}} s_i = S_{j,h}$$

The Lagrangian for this problem is

$$L = \sum_{i \in C_{j,h}} u_i(s_i) - \lambda_{j,h}\left(\sum_{i \in C_{j,h}} s_i - S_{j,h}\right),$$

The first-order condition with respect to $s_i$ is

$$u'_i(s_i) = \lambda_{j,h}, \forall i \in C_{j,h}. \tag{A.1}$$

By Assumption 2, $u_i(s_i)$ is strictly increasing and strictly concave, so $u_i'(\mathrm{s_i}) > 0$ and $u_i''(s_i) < 0$. From the envelope theorem, the marginal revenue of the coalition with respect to its total emission $S_{j,h}$ is equal to the shadow price:

$$U'_{j,h}(S_{j,h}) = \frac{\partial L}{\partial S_{j,h}} = \lambda_{j,h}.$$

Since $u_i'(\mathrm{s_i}) > 0$, we have $U'_{j,h}(S_{j,h}) = \lambda_{j,h} > 0$.

To find the second derivative, differentiate $U'_{j,h}(S_{j,h}) = u'_i(s_i)$ with respect to $S_{j,h}$, where $s_i$ is a function of $S_{j,h}$:

$$U''_{j,h}(S_{j,h}) = u''_i(s_i)\frac{\partial s_i}{\partial S_{j,h}}. \tag{A.2}$$

Equation (A.2) holds for every member $i \in C_{j,h}$. Solving it for $\partial s_i/\partial S_{j,h}$ and summing over all members of the coalition, while using $\sum_{i \in C_{j,h}} \partial s_i/\partial S_{j,h} = 1$, we obtain

$$1 = U''_{j,h}(S_{j,h}) \sum_{i \in C_{j,h}} 1/u''_i(s_i).$$

and hence

$$U''_{j,h}(S_{j,h}) = 1/\sum_{i \in C_{j,h}} 1/u''_i(s_i) < 0,$$

where the inequality follows from $u''_i(s_i) < 0$ for all $i \in C_{j,h}$. Thus $U_{j,h}$ is strictly concave. This completes the proof. □

## Appendix B. Proof of Lemma 2

Suppose, for the sake of contradiction, that the final equilibrium of the cooperative pollution-control game, denoted as State 1, does not achieve the Pareto optimum in pollution control. Let the equilibrium strategies and welfare allocation in State 1 be denoted by $s^1 = (s_1^1, s_2^1, \cdots, s_I^1)$ and $W^1 = (W_1^1, W_2^1, \cdots, W_I^1)$, respectively.

Since State 1 is not Pareto optimal, there exists a feasible joint strategy profile $s^*$ (a Pareto-optimal emission profile) such that the total welfare is strictly greater:

$$W^* = \sum_{i=1}^{I} w_i(s^*, S^*) > \sum_{i=1}^{I} w_i(s^1, S^1) = W^1.$$

Because the game is a Transferable Utility (TU) game, welfare is measured in monetary terms and can be transferred among agents. The grand coalition, consisting of all agents $\{1,2,\cdots,I\}$, can jointly deviate from State 1 and coordinate to choose the Pareto-optimal strategy $s^*$.

Let the total welfare surplus generated by this deviation be $\Delta W = W^* - W^1 > 0$. The grand coalition can design a benefit-transfer scheme such that each agent $i$ receives a welfare increment $\Delta_i = \phi_i \Delta W$, where $\phi_i > 0$ for all $i$ and $\sum_{i=1}^{I} \phi_i = 1$.

Under this new welfare allocation scheme, the final welfare of every agent $i$ becomes:

$$W_i^{new} = W_i^1 + \Delta_i.$$

Since $\Delta_i > 0$ for all $i$, we have $W_i^{new} > W_i^1$ for every agent $i$. Thus, the grand coalition can achieve a strict Pareto improvement over State 1.

As rational welfare maximizers, all agents will agree to form the grand coalition and adopt this new outcome, as it strictly improves their individual welfare. Therefore, State 1 cannot be the final equilibrium of the cooperative game, contradicting the initial assumption.

Hence, the final equilibrium of the cooperative pollution-control game must correspond to a Pareto optimum. □

## Appendix C. Proof of Lemma 3

Let $s^c = (s_1^c, s_2^c, \cdots, s_I^c)$ denote the market equilibrium emission profile and $s^* = (s_1^*, s_2^*, \cdots, s_I^*)$ denote the Pareto-optimal emission profile. We prove the sufficiency and necessity in two steps, following the conditions summarized in Table 1.

Step 1: Sufficiency. Suppose the net price constraint satisfies condition (2$^e$) at the market equilibrium, i.e., $t - T_i = \sum_{\bar{i} \neq i} \frac{\partial d_{\bar{i}}}{\partial S}$. Under the market equilibrium condition (2$^c$), the first-order condition for agent $i$'s welfare maximization is: $t - T_i = \frac{\partial u_i}{\partial s_i} - \frac{\partial d_i}{\partial S}$. Substituting condition (2$^e$) into (2$^c$) yields: $\sum_{\bar{i} \neq i} \frac{\partial d_{\bar{i}}}{\partial S} = \frac{\partial u_i}{\partial s_i} - \frac{\partial d_i}{\partial S}$. Rearranging this equation gives: $\frac{\partial u_i}{\partial s_i} = \frac{\partial d_i}{\partial S} + \sum_{\bar{i} \neq i} \frac{\partial d_{\bar{i}}}{\partial S} = \sum_{\bar{i}=1}^{I} \frac{\partial d_{\bar{i}}}{\partial S}$. This equation is exactly the Pareto-optimality condition (2$^o$). Since the revenue and damage functions are strictly concave and strictly convex (Assumption 2), this solution is unique. Therefore, the market equilibrium achieves the Pareto optimum, i.e., $s^c = s^*$.

Step 2: Necessity. Suppose the market equilibrium achieves the Pareto optimum, meaning $s^c = s^*$. At the Pareto optimum, condition (2$^o$) must hold: $\frac{\partial u_i}{\partial s_i} = \sum_{\bar{i}=1}^{I} \frac{\partial d_{\bar{i}}}{\partial S}$. Simultaneously, at the market equilibrium, condition (2$^c$) must hold: $t - T_i = \frac{\partial u_i}{\partial s_i} - \frac{\partial d_i}{\partial S}$. Substituting the right-hand side of (2$^o$) into (2$^c$) for $\frac{\partial u_i}{\partial s_i}$, we obtain: $t - T_i = (\sum_{j=1}^{I} \frac{\partial d_j}{\partial S}) - \frac{\partial d_i}{\partial S}$. By separating the $i$-th term from the summation, we can rewrite this as: $t - T_i = (\frac{\partial d_i}{\partial S} + \sum_{\bar{i} \neq i} \frac{\partial d_{\bar{i}}}{\partial S}) - \frac{\partial d_i}{\partial S}$. So $t - T_i = \sum_{\bar{i} \neq i} \frac{\partial d_{\bar{i}}}{\partial S}$, which is exactly the price condition (2$^e$). This demonstrates that the price condition is necessary for the market equilibrium to achieve the Pareto optimum.

This completes the proof. □

## Appendix D. Proof of Lemma 4

Let $W^*$ denote the total welfare of all agents at the Pareto optimum. According to Model (6), we have:

$$W^* = \sum_{i=1}^{I} [u_i(s_i^*) - d_i(S^*)]. \quad \text{(D.1)}$$

Let $W_i$ denote the welfare of agent $i$ at the market equilibrium under the price constraints. Based on Model (1), it is given by:

$$W_i = u_i(s_i) - d_i(S) + T_i(\sum_{j=1}^{I} s_j - \sum_{j=1}^{I} \bar{s}_j) + t(\bar{s}_i - s_i) - \bar{t}_i \quad \text{(D.2)}$$

The total welfare of all agents at the market equilibrium is the sum of individual welfare:

$$W = \sum_{i=1}^{I} W_i = \sum_{i=1}^{I} [u_i(s_i) - d_i(S)] + \sum_{i=1}^{I} [T_i(\sum_{j=1}^{I} s_j - \sum_{j=1}^{I} \bar{s}_j) + t(\bar{s}_i - s_i) - \bar{t}_i] \quad \text{(D.3)}$$

By Lemma 3, implementing the price constraints ($2^e$) ensures that the market equilibrium achieves the Pareto optimum. Therefore, the individual emissions and total pollution stock at the market equilibrium are identical to those at the Pareto optimum: $s_i = s_i^*$, and $S = S^*$.

Consequently, the total welfare at the market equilibrium equals the total welfare at the Pareto optimum:

$$W = W^*. \quad \text{(D.4)}$$

Substituting (D.1) and (D.3) into (D.4), we obtain:

$$\sum_{i=1}^{I} [u_i(s_i^*) - d_i(S^*)] + \sum_{i=1}^{I} [T_i(\sum_{j=1}^{I} s_j^* - \sum_{j=1}^{I} \bar{s}_j) + t(\bar{s}_i - s_i^*) - \bar{t}_i] = \sum_{i=1}^{I} [u_i(s_i^*) - d_i(S^*)]$$

Subtracting the direct net benefits $\sum_{i=1}^{I} [u_i(s_i^*) - d_i(S^*)]$ from both sides yields:

$$\sum_{i=1}^{I} [T_i(\sum_{j=1}^{I} s_j^* - \sum_{j=1}^{I} \bar{s}_j) + t(\bar{s}_i - s_i^*) - \bar{t}_i] = 0. \quad \text{(D.5)}$$

Equation (D.5) indicates that the sum of all net transfers (subsidies minus taxes) across all agents is zero. In other words, the total subsidies received by agents must exactly equal the total taxes paid by agents. This guarantees the fiscal balance of the mechanism. ☐

## Appendix E. Proof of Lemma 5

From Lemma 3, the condition for Pareto optimality is given by Eq. (7):

$$t - T_i = \sum_{\bar{i} \neq i} \frac{\partial d_{\bar{i}}}{\partial S}. \quad \text{(E.1)}$$

Summing Eq. (E.1) over all agents $i = 1,2,\cdots,I$, we get $I \cdot t - \sum_{i=1}^{I} T_i = \sum_{i=1}^{I} \sum_{\bar{i} \neq i} \frac{\partial d_{\bar{i}}}{\partial S}$. Since each marginal damage term $\frac{\partial d_{\bar{i}}}{\partial S}$ appears $(I - 1)$ times in the double summation, we have $I \cdot t - \sum_{i=1}^{I} T_i = (I - 1) \sum_{i=1}^{I} \frac{\partial d_i}{\partial S}$. Rearranging this yields the sum of the marginal taxes/subsidies:

$$\sum_{i=1}^{I} T_i = I \cdot t - (I - 1) \sum_{i=1}^{I} \frac{\partial d_i}{\partial S}. \quad \text{(E.2)}$$

From Lemma 4, the fiscal balance condition is given by Eq. (8). We can expand the left-hand side of Eq. (8) using $\bar{s} = \sum_{j=1}^{I} \bar{s}_j$:

$$\sum_{i=1}^{I} [T_i(\sum_{j=1}^{I} s_j - \bar{s}) + t(\bar{s}_i - s_i)] = \sum_{i=1}^{I} \bar{t}_i.$$

Factoring out the terms gives:

$$(\sum_{j=1}^{I} s_j - \bar{s}) \sum_{i=1}^{I} T_i + t \sum_{i=1}^{I} (\bar{s}_i - s_i) = \sum_{i=1}^{I} \bar{t}_i.$$

Since $\bar{s} = \sum_{j=1}^{I} \bar{s}_j$, the second term on the left-hand side simplifies to $t(\bar{s} - \sum_{j=1}^{I} s_j)$. Thus, we obtain:

$$(\sum_{j=1}^{I} s_j - \bar{s})(\sum_{i=1}^{I} T_i - t) = \sum_{i=1}^{I} \bar{t}_i \quad \text{(E.3)}$$

Substituting Eq. (E.2) into Eq. (E.3), we have:

$$(\sum_{j=1}^{I} s_j - \bar{s})[I \cdot t - (I-1) \sum_{i=1}^{I} (\frac{\partial d_i}{\partial S}) - t] = \sum_{i=1}^{I} \bar{t}_i,$$

which simplifies to:

$$\sum_{i=1}^{I} \bar{t}_i = (I-1)(\sum_{j=1}^{I} s_j - \bar{s})(t - \sum_{i=1}^{I} \frac{\partial d_i}{\partial S}). \quad \text{(E.4)}$$

Equation (E.4) characterizes the solution space for the price constraints. We can satisfy this condition in two canonical ways:

Solution I (Price Constraint I):

Suppose the aggregate actual emissions differ from the aggregate initial permits ($\sum_{j=1}^{I} s_j \neq \bar{s}$). Assuming zero lump-sum transfers ($\bar{t}_i = 0$ for all $i$), the left-hand side of Eq. (E.4) equals zero. Since $I > 1$ and the deviation of total emissions is non-zero, the remaining factor must be zero:$t - \sum_{i=1}^{I} \frac{\partial d_i}{\partial S} = 0$. Rearranging this yields the uniform emission price: $t = \sum_{i=1}^{I} \frac{\partial d_i}{\partial S}$. Substituting Eq. (9) back into Eq. (E.1) yields the individual marginal tax/subsidy: $T_i = \frac{\partial d_i}{\partial S}$. This constitutes Price Constraint I (Equations 9 and 10 in Table 2), and it naturally satisfies the fiscal balance condition $\bar{t}_i = 0$ (Equation 12).

Solution II (Price Constraint II):

Alternatively, we can set the uniform emission price to zero ($t = 0$). To satisfy Eq. (E.1), the individual marginal tax/subsidy must be $T_i =- \sum_{\bar{i} \neq i} \frac{\partial d_{\bar{i}}}{\partial S}$. Under this setting, if $\sum_{j=1}^{I} s_j \neq \bar{s}$, Eq. (E.4) implies that non-zero aggregate lump-sum transfers $\bar{t}_i$ are required to maintain fiscal balance:

$$\sum_{i=1}^{I}\bar{t}_i =-(I-1)(\sum_{j=1}^{I}s_j-\bar{s})\sum_{i=1}^{I}\frac{\partial d_i}{\partial S}.$$

This provides a canonical implementation of Price Constraint II.

Therefore, Lemma 5 is proven: Pareto optimality (Eq. 7) and fiscal balance (Eq. 8) together restrict the price constraints $(t, T_i, \bar{t}_i)$ to the optimal equilibrium prices, which can be implemented via either Price Constraint I or Price Constraint II. ☐

## Appendix F. Proof of Lemma 7

Under Price Constraint I, the market equilibrium achieves Pareto optimality (Lemma 3). Therefore, the actual emissions and total pollution stock are equal to their Pareto-optimal levels, i.e., $s_i = s_i^*$ and $S = S^*$. The welfare of agent $i$ under Price Constraint I is given by:

$$W_i = u_i(s_i^*) - d_i(S^*) + T_i^*(\sum_{j=1}^{I}s_j^* - \bar{s}) + t^*(\bar{s}_i - s_i^*), \tag{F.1}$$

where $\bar{s} = \sum_{j=1}^{I}\bar{s}_j$ is the total initial permit endowment, and $t^* = \sum_{j=1}^{I}d'_j(S^*)$ and $T_i^* = d'_i(S^*)$ are the optimal price constraints.

We prove the lemma in three steps.

Step 1: The total welfare is independent of the initial allocation.

Summing Eq. (F.1) over all agents $i = 1,2,\cdots,I$ yields:

$$\sum_{i=1}^{I}W_i = \sum_{i=1}^{I}[u_i(s_i^*) - d_i(S^*)] + (\sum_{j=1}^{I}s_j^* - \bar{s})\sum_{i=1}^{I}T_i^* + t^*\sum_{i=1}^{I}(\bar{s}_i - s_i^*).$$

Since $\bar{s} = \sum_{i=1}^{I}\bar{s}_i$ and $t^* = \sum_{i=1}^{I}T_i^*$, the equation simplifies to:

$$\sum_{i=1}^{I}W_i = W^* + (\sum_{j=1}^{I}s_j^* - \bar{s})t^* + t^*(\bar{s} - \sum_{j=1}^{I}s_j^*) = W^*.$$

Thus, the final equilibrium total welfare equals the total welfare at the Pareto optimum, regardless of the initial allocation of permits.

Step 2: Any Pareto-optimal welfare distribution can be achieved.

Suppose we target an arbitrary welfare distribution $(W_1, W_2, \cdots, W_I)$ such that $\sum_{i=1}^{I}W_i = W^*$. Let $b_i = u_i(s_i^*) - d_i(S^*)$ denote the direct net benefit of agent $i$ at the Pareto optimum.

Rearranging Eq. (F.1) allows us to solve for the required initial permit allocation $\bar{s}_i$ that realizes the target welfare $W_i$:

$$t^*\bar{s}_i - T_i^*\bar{s} = W_i - b_i - T_i^*\sum_{j=1}^{I}s_j^* + t^*s_i^*.$$

Let the right-hand side be a constant $A_i$. The required allocation can then be expressed as:

$$\bar{s}_i = \frac{A_i}{t^*} + \frac{T_i^*}{t^*}\bar{s}. \tag{F.2}$$

Since $\sum_{i=1}^{I} W_i = W^* = \sum_{i=1}^{I} b_i$ and $t^* = \sum_{i=1}^{I} T_i^*$, it immediately follows that:

$$\sum_{i=1}^{I} A_i = W^* - W^* - t^* \sum_{j=1}^{I} s_j^* + t^* \sum_{i=1}^{I} s_i^* = 0.$$

Summing Eq. (F.2) over all $i$ yields $\sum_{i=1}^{I} \bar{s}_i = \bar{s}$, which confirms that the required individual allocations exactly satisfy the aggregate permit constraint. Equation (F.2) therefore defines a one-dimensional family of solutions parameterized by $\bar{s}$. By choosing a sufficiently large $\bar{s}$, we can ensure the required $\bar{s}_i \geq 0$ for all $i$. Hence, any Pareto-optimal distribution can be realized by designing a specific initial permit allocation scheme.

Step 3: The welfare effects of permit endowments.

Taking the partial derivative of $W_i$ in Eq. (F.1) with respect to agent $i$'s own endowment $\bar{s}_i$ (while keeping others' endowments fixed, so $\partial\bar{s}/\partial\bar{s}_i = 1$), we get:

$$\frac{\partial W_i}{\partial \bar{s}_i} = t^* - T_i^* = \sum_{j \neq i} \frac{\partial d_j}{\partial S^*} > 0. \tag{F.3}$$

This shows that an agent's welfare is directly proportional to its own initial permits.

Similarly, taking the partial derivative with respect to another agent's endowment $\bar{s}_j$ $(j \neq i)$, we get:

$$\frac{\partial W_i}{\partial \bar{s}_j} =- T_i^* =- \frac{\partial d_i}{\partial S^*} < 0. \tag{F.4}$$

This shows that an agent's welfare is inversely proportional to the total permits held by other agents.

This completes the proof. □

## Appendix G. Proof of Lemma 8

Suppose under Coalition Structure 1, all agents adopt non-cooperative strategies. Under Coalition Structure 2, certain agents adopt cooperative strategies. Let $s_{i,1}$ and $s_{i,2}$ denote the individual emissions of agent $i$ under Coalition Structures 1 and 2, respectively. Let $S_1$ and $S_2$ denote the total pollution stocks under the two structures. For simplicity, assume that under Structure 2, coalitions $j = 1,2,\cdots,\bar{J}$ adopt non-cooperative strategies, while the remaining coalitions $j = \bar{J}+1, \bar{J}+2, \cdots, J_2$ adopt cooperative strategies.

The equilibrium conditions for agent $i$ under Coalition Structure 1 (all non-cooperative) are:

$$\frac{\partial u_i(s_{i,1})}{\partial s_{i,1}} = \frac{\partial d_i(S_1)}{\partial S_1}, \forall i. \tag{G.1}$$

The equilibrium conditions under Coalition Structure 2 are:

$$\frac{\partial u_i(s_{i,2})}{\partial s_{i,2}} = \frac{\partial d_i(S_2)}{\partial S_2}, \text{ for } j = 1,2,\cdots,\bar{J}, \tag{G.2}$$

$$\frac{\partial u_i(s_{i,2})}{\partial s_{i,2}} = \sum_{k\in C_{j,2}} \frac{\partial d_k(S_2)}{\partial S_2}, \text{ for } j = \bar{J}+1, \bar{J}+2, \cdots, J_2. \tag{G.3}$$

Step 1: Total emissions decrease ($S_1 > S_2$).

We prove $S_1 > S_2$ by contradiction. Suppose $S_1 \leq S_2$. Because damage functions are strictly convex ($d''_i > 0$), we have:

$$\frac{\partial d_i(S_1)}{\partial S_1} \leq \frac{\partial d_i(S_2)}{\partial S_2}. \tag{G.4}$$

For agents in non-cooperative coalitions ($j \leq \bar{J}$), from (G.1) and (G.2), we get:

$$\frac{\partial u_i(s_{i,1})}{\partial s_{i,1}} = \frac{\partial d_i(S_1)}{\partial S_1} \leq \frac{\partial d_i(S_2)}{\partial S_2} = \frac{\partial u_i(s_{i,2})}{\partial s_{i,2}}. \tag{G.5}$$

Since $u_i$ is strictly concave ($u''_i < 0$), this implies $s_{i,1} \geq s_{i,2}$.

For agents in cooperative coalitions ($j > \bar{J}$), because the coalition internalizes the externality, $\sum_{k\in C_{j,2}} \frac{\partial d_k(S_2)}{\partial S_2} > \frac{\partial d_i(S_2)}{\partial S_2}$. From (G.1), (G.3), and (G.4), we have:

$$\frac{\partial u_i(s_{i,1})}{\partial s_{i,1}} = \frac{\partial d_i(S_1)}{\partial S_1} \leq \frac{\partial d_i(S_2)}{\partial S_2} < \sum_{k\in C_{j,2}} \frac{\partial d_k(S_2)}{\partial S_2} = \frac{\partial u_i(s_{i,2})}{\partial s_{i,2}}. \tag{G.6}$$

This implies $s_{i,1} > s_{i,2}$.

Summing $s_{i,1} > s_{i,2}$ over all agents (with strict inequality for cooperative agents), we obtain $S_1 > S_2$, which contradicts the assumption $S_1 \leq S_2$. Therefore, $S_1 > S_2$.

Step 2: Non-cooperative agents increase emissions and welfare.

Since $S_1 > S_2$, we have $\frac{\partial d_i(S_1)}{\partial S_1} > \frac{\partial d_i(S_2)}{\partial S_2}$. For non-cooperative agents ($j \leq \bar{J}$), combining (G.1) and (G.2) yields $\frac{\partial u_i(s_{i,1})}{\partial s_{i,1}} > \frac{\partial u_i(s_{i,2})}{\partial s_{i,2}}$, which implies $s_{i,1} < s_{i,2}$. Thus, their emissions increase.

Their welfare change is given by:

$$[u_i(s_{i,2}) - d_i(S_2)] - [u_i(s_{i,1}) - d_i(S_1)].$$

Given $s_{i,2} > s_{i,1}$ and the strict monotonicity of the revenue function ($u'_i > 0$), the revenue increases, i.e., $u_i(s_{i,2}) > u_i(s_{i,1})$. Meanwhile, because $S_2 < S_1$ and $d'_i > 0$, the damage decreases, yielding $d_i(S_2) < d_i(S_1)$. Since both effects are strictly positive, their welfare strictly increases.

Step 3: Cooperative agents decrease emissions, but welfare change is ambiguous.

For cooperative agents ($j > \bar{J}$), they reduce their emissions, i.e., $s_{i,1} > s_{i,2}$. Their welfare change is:

$$\Delta W_i = [u_i(s_{i,2}) - u_i(s_{i,1})] - [d_i(S_2) - d_i(S_1)].$$

The decrease in emissions ($s_{i,2} < s_{i,1}$) leads to a revenue loss, making the first bracketed term negative. In contrast, the reduction in total pollution ($S_2 < S_1$) lowers the damage, making the second bracketed term negative and thus its contribution to the welfare change positive. The net welfare effect thus depends on the relative magnitudes of the revenue loss versus the damage reduction; therefore, the welfare change for cooperative agents remains ambiguous.

This completes the proof. □

## Appendix H. Proof of Lemma 9

(i) Under the *w*-IPA scheme, the initial permit allocation is set to the Nash equilibrium emissions, i.e., $\bar{s}_i = \tilde{s}_i$ for all $i$. Combined with Price Constraint I, the uniform price is $t = \sum_{j=1}^{I} d'_j(S)$ and the individual marginal tax/subsidy is $T_i = d'_i(S)$.

Given these price constraints, the welfare maximization problem for agent $i$ (from Model 1) is:

$$max: u_i(s_i) - d_i(S) + T_i\left(\sum_{j=1}^{I} s_j - \sum_{j=1}^{I} \bar{s}_j\right) + t(\bar{s}_i - s_i) - \bar{t}_i.$$

Taking the first-order condition with respect to $s_i$ (noting that $\frac{\partial S}{\partial s_i} = 1$), we obtain:

$$u'_i(s_i) - d'_i(S) + T_i - t = 0.$$

Substituting the price constraints $t = \sum_{j=1}^{I} d'_j(S)$ and $T_i = d'_i(S)$ into the condition yields:

$$u'_i(s_i) = \sum_{j=1}^{I} d'_j(S), \forall i \in \{1,2,\cdots,I\}. \tag{H.1}$$

(Note that the initial allocation $\bar{s}_i = \tilde{s}_i$ enters the welfare function as a constant transfer $t\tilde{s}_i$ and $T_i \sum_{j=1}^{I} \tilde{s}_j$. Its derivative with respect to $s_i$ is zero, so it does not distort the marginal incentives.)

Equation (H.1) is mathematically identical to the first-order condition for the Pareto-optimality Model (6), which is $u'_i(s_i^*) = \sum_{j=1}^{I} d'_j(S^*)$. Because the revenue functions are strictly concave ($u''_i < 0$) and the damage functions are strictly convex ($d''_i > 0$), this system of non-linear equations has a unique solution. Therefore, the market equilibrium induced by the *w*-IPA scheme must coincide with the Pareto optimum, yielding $s_i = s_i^*$ for all $i \in \{1,2,\cdots,I\}$.

(ii) To prove that $\sum_{i\in C} W_{i,w-value} \geq v_\gamma(C)$ for any deviating coalition $C \subseteq \{1,2,\cdots,I\}$, we establish an upper bound for the deviation value and compare it with the *w-value* allocation.

If coalition $C$ deviates, the outsiders $j \notin C$ choose their Nash best responses $s_j^{BR}$. Let $s_k^{dev}$ denote the optimal emissions chosen by coalition members $k \in C$. Let $S^{BR} = S_0 + \sum_{k \in C} s_k^{dev} + \sum_{j \notin C} s_j^{BR}$ be the resulting total pollution stock. For notational uniformity, we write $s_j^{dev} = s_j^{BR}$ for the outsiders $j \notin C$. The deviation value is:

$$v_\gamma(C) = \sum_{k \in C} [u_k(s_k^{dev}) - d_k(S^{BR})]. \tag{H.2}$$

Step 1: Establish an upper bound for $v_\gamma(C)$.

Since $u_k$ is strictly concave and $d_k$ is strictly convex, their first-order Taylor expansions around the Pareto optimum $(s^*, S^*)$ yield the following inequalities:

$$u_k(s_k^{dev}) \le u_k(s_k^*) + u'_k(s_k^*)(s_k^{dev} - s_k^*),$$

$$d_k(S^{BR}) \ge d_k(S^*) + d'_k(S^*)(S^{BR} - S^*).$$

Substituting these into Eq. (H.2) and using the Pareto-optimality condition $u'_k(s_k^*) = \sum_{m=1}^{I} d'_m(S^*) = t^*$, we obtain:

$$v_\gamma(C) \le \sum_{k \in C} [u_k(s_k^*) - d_k(S^*)] + t^* \sum_{k \in C} (s_k^{dev} - s_k^*) - T_C^*(S^{BR} - S^*), \tag{H.3}$$

where $T_C^* = \sum_{k \in C} d'_k(S^*)$.

Step 2: Express the *w*-value allocation.

From Definition 6, the total *w*-value for coalition $C$ is given by:

$$\sum_{k \in C} W_{k,w-value} = \sum_{k \in C} [u_k(s_k^*) - d_k(S^*)] + t^* \sum_{k \in C} (\tilde{s}_k - s_k^*) - T_C^* \sum_{m=1}^{I} (\tilde{s}_m - s_m^*), \tag{H.4}$$

where $\tilde{s}_m$ is the fully non-cooperative Nash equilibrium emission.

To compare (H.3) and (H.4), we take the difference between the transfer components:

$$\sum_{k \in C} W_{k,w-value} - v_\gamma(C) \ \ge t^* \sum_{k \in C} (\tilde{s}_k - s_k^{dev}) - T_C^* \sum_{m=1}^{I} (\tilde{s}_m - s_m^*) + T_C^*(S^{BR} - S^*). \tag{H.5}$$

Since $S^{BR} - S^* = \sum_{m=1}^{I} (s_m^{dev} - s_m^*)$, the last two terms in (H.5) can be combined:

$$- T_C^* \sum_{m=1}^{I} (\tilde{s}_m - s_m^*) + T_C^* \sum_{m=1}^{I} (s_m^{dev} - s_m^*) =- T_C^* \sum_{m=1}^{I} (\tilde{s}_m - s_m^{dev}).$$

Substituting this back into (H.5) yields:

$$\sum_{k \in C} W_{k,w-value} - v_\gamma(C) \ \ge t^* \sum_{k \in C} (\tilde{s}_k - s_k^{dev}) - T_C^* \sum_{m=1}^{I} (\tilde{s}_m - s_m^{dev}). \tag{H.6}$$

Step 3: Prove that the inequality holds.

We can rewrite (H.6) by splitting the aggregate summation into members of $C$ and outsiders $j \notin C$:

$$\sum_{k \in C} W_{k,w-value} - v_\gamma(C) \ \ge (t^* - T_C^*) \sum_{k \in C} (\tilde{s}_k - s_k^{dev}) - T_C^* \sum_{j \notin C} (\tilde{s}_j - s_j^{dev}). \tag{H.7}$$

Note that $t^* = \sum_{m=1}^{I} d'_m(S^*)$. Therefore, $(t^* - T_C^*) = \sum_{j \notin C} d'_j(S^*)$, which represents the total marginal damage of the outsiders.

We now verify that the right-hand side of (H.7) is nonnegative. By Lemma 8, when a subset of agents adopts cooperative strategies, the equilibrium emissions of these cooperative agents decrease relative to the fully non-cooperative Nash equilibrium, whereas the emissions of the non-cooperative agents increase. Hence, for the deviation considered here:

$$\tilde{s}_k \geq s_k^{dev}, \forall k \in C \text{ and } \tilde{s}_j \leq s_j^{dev}, \forall j \notin C.$$

Consequently, the first term on the right-hand side of (H.7) is nonnegative, since $t^* - T_C^* > 0$ and $\tilde{s}_k - s_k^{dev} \geq 0$ for all $k \in C$; the second term is nonnegative as well, since $T_C^* > 0$ and $\tilde{s}_j - s_j^{dev} \leq 0$ for all $j \notin C$.

$$(t^* - T_C^*) \sum_{k \in C} (\tilde{s}_k - s_k^{dev}) - T_C^* \sum_{j \notin C} (\tilde{s}_j - s_j^{dev}) \geq 0.$$

Therefore, Eq. (H.7) implies:

$$\sum_{i \in C} W_{i,w-value} \geq v_\gamma(C), \forall C \subseteq \{1,2,\cdots,I\}.$$

This establishes that the *w*-value allocation cannot be blocked by any coalition, meaning it belongs to the $\gamma$-core. ☐

## Appendix I. Proof of Lemma 10

Lemma 10 states that among all Category I welfare distribution schemes, only the *w*-IPA welfare distribution can serve as the equilibrium of the cooperative game. We prove this by showing that any incomplete-cooperative coalition structure will eventually unravel.

Let $P^P$ denote an arbitrary incomplete-cooperative coalition structure, and $P^N$ denote the fully non-cooperative coalition structure. The Nash equilibrium emissions of agent $i$ under $P^P$ and $P^N$ are denoted by $\tilde{s}_i^P$ and $\tilde{s}_i^N$, respectively. Under a Category I scheme, the initial permit allocation is based on the Nash equilibrium emissions, so $\bar{s}_i^P = \tilde{s}_i^P$, and $\bar{s}_i^N = \tilde{s}_i^N$.

Consider a coalition $C$ in $P^P$ comprising $I_C > 1$ members. If at least one member of $C$ adopts a non-cooperative strategy, $P^P$ dissolves toward $P^N$. Under $P^P$, the members of $C$ internalize the externality by equating their marginal revenue to the coalition's aggregate marginal damage, $u'_i(s_i) = \sum_{k \in C} d'_k(S)$. When they deviate to $P^N$, they only care about their own damage. By Lemma 8, this abandonment of collective responsibility strictly increases the Nash equilibrium emissions of the members of $C$, while strictly reducing those of the outsiders $j \notin C$. Consequently:

$$\bar{s}_C^N - \bar{s}_C^P > 0,\ \sum_{j \notin C} (\bar{s}_j^N - \bar{s}_j^P) < 0, \tag{I.1}$$

where $\bar{s}_C^k = \sum_{i \in C} \bar{s}_i^k$ and $\bar{s}^k = \sum_{i=1}^I \bar{s}_i^k$ for $k \in \{p, N\}$.

Given Price Constraint I, the total welfare of coalition $C$ based on the initial allocation $\bar{s}^k$ is:

$$W_C^k = \sum_{i \in C} [u_i(s_i^*) - d_i(S^*)] + t^* \bar{s}_C^k - t^* \sum_{i \in C} s_i^* + T_C^* \sum_{j=1}^I s_j^* - T_C^* \bar{s}^k \tag{I.2}$$

where $T_C^* = \sum_{i \in C} T_i^*$ and $t^* = \sum_{i=1}^I T_i^*$. We compare the welfare of coalition $C$ under $P^P$ and $P^N$:

$$W_C^N - W_C^P = t^* (\bar{s}_C^N - \bar{s}_C^P) - T_C^* (\bar{s}^N - \bar{s}^P). \tag{I.3}$$

Substitute $t^* = T_C^* + \sum_{j \notin C} T_j^*$ into Eq. (I.3), yielding:

$$W_C^N - W_C^P = (\sum_{j \notin C} T_j^*)(\bar{s}_C^N - \bar{s}_C^P) - T_C^* \sum_{j \notin C} (\bar{s}_j^N - \bar{s}_j^P). \tag{I.4}$$

Substituting the sign results in (I.1) into Eq. (I.4), and noting that each transfer rate $T_j^*$ is strictly positive, both terms on the right-hand side of Eq. (I.4) are strictly positive. Therefore:

$$W_C^N - W_C^P > 0. \tag{I.5}$$

Equation (I.5) indicates that members of $C$ strictly prefer the permit allocation based on $P^N$. As rational welfare maximizers, they will withdraw from $P^P$ to induce $P^N$. By backward induction, any initial allocation derived from an incomplete-cooperative Nash equilibrium will be abandoned in favor of the allocation based on the fully non-cooperative Nash equilibrium—namely, the *w*-IPA welfare distribution. Once the *w*-IPA is in place, Lemma 9 shows that no coalition can obtain a higher welfare by any further deviation, because the *w-value* allocation satisfies the $\gamma$-core condition. Therefore, only the *w*-IPA welfare distribution can constitute the final equilibrium of the cooperative pollution-control game. □